\documentclass[
showkeys,	%	Control display of Keywords: line.
reprint,	%	Two-column formatting.
amsmath,	%	Load amsmath package.- AMS-LATEX features.
amssymb,	%	Load amssymb package.-  AMS symbols.
amsfonts,	%	Load amsfonts package.-  AMS font support.
aps,		%	American Physical Society styling.
pre,%
groupedaddress,% Group authors with same affiliations.
superscriptaddress,% Associate authors with 			affiliations via superscript numbers. Appropriate for collaborations or if several authors share some, but not all, affiliations.
a4paper,%
nofootinbib,%	Control location of footnotes.
eprint,		%	arXiv e-print identifiers in bibliography.
final,		%	Don’t mark overfull lines.
preprintnumbers, % Control display of preprint numbers given by \preprint command.
]{revtex4-2}

\usepackage[tbtags]{mathtools}
\usepackage{amsmath,amsfonts,mathdots,amssymb,yfonts,calc}
\usepackage{graphicx,graphics,xcolor}
\usepackage{subfigure}
\usepackage{tabularx}
\usepackage{booktabs}
\usepackage{siunitx}
\usepackage{float}
\usepackage{tabularray}
\usepackage{multirow}
\usepackage{physics}
\usepackage{dcolumn}% Align table columns on decimal point
\usepackage[utf8]{inputenc}  % <- ADD THIS FIRST
\usepackage[T1]{fontenc} 
\usepackage{natbib}
\usepackage{booktabs}
\usepackage{hyperref}% Load hyperref LAST
\hypersetup{
    colorlinks=true,
    linkcolor=blue,
    filecolor=magenta,      
    urlcolor=blue!80,
    citecolor=cyan,
    linktocpage=true,
    breaklinks=false,
}

\begin{document}
\title{Universality classes of chaos in non-Markovian dynamics}
%*******************************************************
\author{Vinesh Vijayan}
\email[]{vinesh.physics@rathinam.in}
\affiliation{Department of Science and Humanities, Rathinam Technical Campus, Coimbatore, India -641021}

%**********************************************************

\date{\today}

\begin{abstract}
Classical chaos theory rests on the notion of universality, whereby disparate dynamical systems share identical scaling laws. Existing universality classes, however, implicitly assume Markovian dynamics. Here, a logistic map endowed with power-law memory is used to show that Feigenbaum universality breaks down when temporal correlations decay sufficiently slowly. A critical memory exponent is identified that separates perturbative and memory‑dominated regimes, demonstrating that long-range memory acts as a relevant renormalisation operator and generates a new universality class of chaotic dynamics. The onset of chaos is accompanied by fractional scaling of Lyapunov exponents, in quantitative agreement with analytical predictions. These results establish temporal correlations as a previously unexplored axis of universality in chaotic systems, with implications for physical, biological and geophysical settings where memory effects are intrinsic.
\end{abstract}

\keywords{Chaos; critical scaling; Non Markovian; Universality}
%\maketitle must follow title, authors, abstract, and keyword
\maketitle
%\tableofcontents
\pagestyle{plain}
\section{Introduction}
Deterministic low-dimensional iterative maps provide a paradigmatic setting for universality, exemplified by the Feigenbaum period-doubling cascade in the logistic family and related one-dimensional maps\cite{Feigenbaum1978, Feigenbaum1979, Cvitanovic1987, May1976, Adler1965, Kuzmin1928, Arnold1961, Pomeau1980}. In all such systems, the future state depends only on the present state, placing them within a Markovian framework. The universal scaling laws are governed by a finite-dimensional renormalisation group fixed point, with Feigenbaum constants $\delta_F \sim 4.669$ and $\alpha_F \sim 2.503$ characterising the geometric accumulation of bifurcation points and the scaling and slope of the phase-space structure, respectively\cite{Feigenbaum1978, Feigenbaum1979, Cvitanovic1987, Grebogi1983}. This universality extends to a broad class of smooth, unimodal maps and underlies the route to chaos in diverse physical systems, ranging from driven oscillators and fluid convection to electronic circuits\cite{Linsay1981, Gollub1980, Libchaber1982, Gollub1975, Testa1982, Linsay1983}.\\
In real systems, the dynamics depend on an extended temporal history rather than a single state, the presence of a long‑lived memory, from viscoelastic and glassy materials to biological feedback networks, delay‑coupled oscillators and non‑Markovian quantum environments\cite{Parlitz1987, Longtin1993, Paz1993, Wolf2011, Walkama2020, Chaudhary2021, Karmakar2020, Lerner2022, Goyal2022, Koseska2023}. In such non‑Markovian set‑ups, the standard renormalization group picture of chaos breaks down, and the impact of memory on routes to chaos and universality remains poorly understood, other than some numerical case studies\cite{Watts1998, Munoz2007, Restrepo2007, Li2023, Pawela2025, Borah2022, Redner2025}. For example, in non‑Markovian quantum systems, memory effects can modify decoherence and thermalization, and their roles in generating new classes of chaos are largely underexplored. A clean universality classification analogous to the Feigenbaum scenario is lacking in systems with time delays, where memory induces complex bifurcation structures and alters the stability of the periodic orbits. Long‑range temporal correlations in complex systems, such as in neural networks, ecological models and certain driven dissipative media, point towards a fundamental reshaping of the deterministic chaos landscape\cite{Mackey1977, Pikovsky2005}.\\
A general classification of universality in chaos remains elusive, despite the known role of non-Markovian effects in modifying chaotic dynamics in specific systems such as delay-coupled oscillators and optomechanical devices. Here, we construct a minimal, analytically tractable non-Markovian chaotic system by endowing the logistic map with a power-law memory kernel, and show that memory induces new universality classes governed by the memory exponent, sharply distinguishing perturbative and non-perturbative regimes. The results identify distinct universality classes of chaos in non-Markovian dynamics, providing a minimal theoretical framework for memory-induced transitions in complex systems ranging from delayed feedback oscillators to long-range interacting media.
\section{The Minimal Model}
\subsection{Non-Markovian Logistic Map with Power Law memory}
Consider the non-Markovian logistic map, which unifies chaotic dynamics with nonlocal temporal dependencies. The iteration is defined by
\begin{equation}
x_{n+1} = r x_n (1 - x_n) + \epsilon \sum_{k=1}^n k^{-\alpha} x_{n-k}, \quad x_n \in [0,1]
\label{eq1}
\end{equation}
Here, $r \in [0,4]$ is the logistic growth parameter, $\epsilon \ll 1$ the memory strength, and $\alpha > 0$ the memory exponent ($\alpha > 1$ for short memory, $\alpha \leq 1$ for long memory). In the limit $\epsilon \to 0$, the map reduces to the classical logistic map; finite $\epsilon$ introduces nonzero temporal correlations.
The first term on the right-hand side is the classical logistic growth term, a hallmark system for studying chaos. The memory term explicitly weights all prior states $x_{n-k}$ with the power-law kernel $k^{-\alpha}$, introducing temporal nonlocality whereby future evolution correlates with the distant past, requiring an infinite-dimensional state for complete description. Physically, these correlations induce information flows, altered decay rates, and polylogarithmic characteristic equations, in stark contrast to the Markovian case.
\subsubsection{Linear Stability Analysis}
The stability of the recurrence relation in Equation~(\ref{eq1}) depends on $r$, $\epsilon$ and $\alpha$. Linearizing the infinite-history dynamics, the fixed points satisfy
\begin{equation}
x^* = f(x^*) + \epsilon x^* \sum_{k=1}^\infty k^{-\alpha}
\label{eq2}
\end{equation}
This assumes that, after initial transients, all past states equal the constant $x_{n-k} = x^*$; in the limit $n \to \infty$, Equation~(\ref{eq2}) follows. The trivial fixed point of the system is $x^* = 0$, and the non-trivial fixed points solve
\begin{equation}
x^* = r x^* (1 - x^*) + \epsilon x^* \zeta(\alpha),
\label{eq3}
\end{equation}
where $\zeta(\alpha) = \sum_{k=1}^\infty k^{-\alpha}$ is the Riemann zeta function, convergent for $\alpha > 1$.

Assume $x_n = x^* + \delta x_n$ with small perturbations $\delta x_n$; then equation~(\ref{eq:map}) becomes
\begin{equation}
x^* + \delta x_{n+1} = f(x^* + \delta x_n) + \epsilon \sum_{k=1}^n k^{-\alpha} (x^* + \delta x_{n-k}).
\label{eq4}
\end{equation}
Taylor expanding $f$ around $x^*$,
\begin{equation*}
f(x^* + \delta x_n) = f(x^*) + f'(x^*)\, \delta x_n + \mathcal{O}\bigl((\delta x_n)^2\bigr).
\end{equation*}
Splitting the sum on the right-hand side,
\begin{equation*}
\epsilon \sum_{k=1}^n k^{-\alpha} (x^* + \delta x_{n-k}) = \epsilon x^* \sum_{k=1}^n k^{-\alpha} + \epsilon \sum_{k=1}^n k^{-\alpha} \delta x_{n-k}.
\end{equation*}
Subtracting the fixed-point condition and keeping only linear terms in $\delta x_n$, we obtain
\begin{equation*}
\begin{split}
\delta x_{n+1} &= f'(x^*)\, \delta x_n + \epsilon \sum_{k=1}^n k^{-\alpha} \delta x_{n-k} \\
               &\quad + \epsilon x^* \left( \sum_{k=1}^n k^{-\alpha} - \sum_{k=1}^\infty k^{-\alpha} \right) + \mathcal{O}\bigl((\delta x_n)^2\bigr).
\end{split}
\end{equation*}
For large $n$, the tail $\sum_{k=n+1}^{\infty} k^{-\alpha} \to 0$ so that $\sum_{k=1}^{\infty} k^{-\alpha} \simeq \zeta(\alpha)$. Neglecting higher-order terms, this yields the non-Markovian linearized equation
\begin{equation}
\delta x_{n+1} = f'(x^*)\, \delta x_n + \epsilon \sum_{k=1}^{\infty} k^{-\alpha} \delta x_{n-k}.
\label{eq5}
\end{equation}
Assuming asymptotic exponential behaviour $\delta x_n \sim \lambda^n$ as $n \to \infty$, with $\lambda \in \mathbb{C}$ and $|\lambda| \sim 1$ near marginal stability, substitution into the linearized equation yields
\begin{equation*}
\lambda^{n+1} = a\, \lambda^n + \epsilon \sum_{k=1}^n k^{-\alpha} \lambda^{n-k}.
\end{equation*}
Dividing by $\lambda^n$ and taking $n \to \infty$ gives
\begin{equation*}
\lambda = a + \epsilon \sum_{k=1}^{\infty} \frac{\lambda^{-k}}{k^\alpha},
\end{equation*}
where $a = f'(x^*)$. Defining the polylogarithm
\begin{equation}
\mathrm{Li}_\alpha(z) = \sum_{k=1}^{\infty} \frac{z^k}{k^\alpha},
\label{eq6}
\end{equation}
the characteristic equation becomes
\begin{equation}
\lambda = a + \epsilon\, \mathrm{Li}_\alpha(\lambda^{-1}).
\label{eq7}
\end{equation}
This is the characteristic equation, the central analytical object of the paper. The fixed point is asymptotically stable if all roots $\lambda$ satisfy $|\lambda| < 1$. For small $\epsilon$, the roots are perturbed from the memoryless case, with the memory term $\sum_{k=1}^{\infty} k^{-\alpha} \lambda^{-k}$ acting as a nonlocal multiplier, analytic in $|\lambda| > 1$ for $\alpha > 0$.
\begin{figure*}[t]
  \centering
  \includegraphics[width=0.7\textwidth]{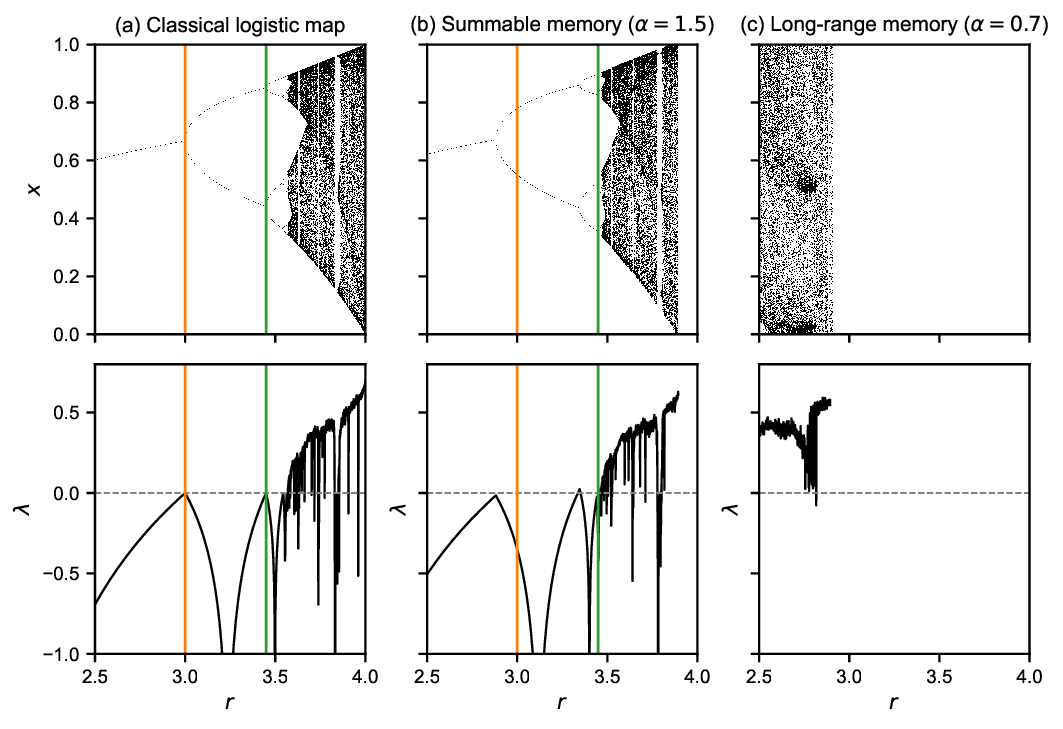}
  \caption{
    \textbf{Bifurcation diagrams and Lyapunov exponents.}(a) Classical logistic map ($\epsilon=0$).
(b) Summable memory ($\epsilon=0.02$,~$\alpha=1.5$).
(c) Long-range memory ($\epsilon=0.02$,~$\alpha=0.7$).
Top row: bifurcation diagram; bottom row: Lyapunov exponent~$\lambda(r)$. Vertical lines mark the first ($r_1=3.0$) and second ($r_2\approx 3.45$) period-doubling bifurcations of the classical map; the dashed line at~$\lambda=0$ marks the onset of chaos.    
  }
  \label{fig1}
\end{figure*}

In FIG.~(\ref{fig1}), the three panels demonstrate that temporal memory acts as an independent control parameter for chaotic dynamics. Summable memory preserves classical universality, albeit with a compressed bifurcation structure and a shift in critical parameter values, as indicated by the vertical lines marking the period‑doubling cascade. In contrast, long‑range memory fundamentally changes the balance between local instability and global feedback. Chaos becomes non‑monotonic in the control parameter, emerges without a Feigenbaum cascade, and is ultimately suppressed at large driving. The combined bifurcation–Lyapunov analysis shows that these effects are not numerical artifacts, but reflect genuine non‑Markovian dynamics. This figure provides direct numerical evidence that long‑range temporal correlations define a new universality class of chaos, characterized by early onset, fractional critical scaling, and re‑entrant stabilization.

\subsubsection{Case I: Summable Memory ($\alpha >1$)}
For $|\lambda| \ge 1$, we have $|\lambda^{-1}| \le 1$. The series then satisfies
\begin{equation}
\bigl|\mathrm{Li}_\alpha(\lambda^{-1})\bigr| \le \sum_{k=1}^\infty \frac{|\lambda^{-1}|^k}{k^\alpha} \le \sum_{k=1}^\infty \frac{1}{k^\alpha} = \zeta(\alpha) < \infty,
\label{eq8}
\end{equation}
so that $\mathrm{Li}_\alpha(\lambda^{-1})$ is bounded and analytic in the region $|\lambda| \ge 1$ for $\alpha > 1$.
Taking absolute values of equation~(\ref{eq7}) yields
\begin{equation}
|\lambda| \le |a| + \epsilon \bigl|\mathrm{Li}_\alpha(\lambda^{-1})\bigr| \le |a| + \epsilon\, \zeta(\alpha).
\label{eq9}
\end{equation}
Thus, if $|a| + \epsilon\, \zeta(\alpha) < 1$, no roots can satisfy $|\lambda| \ge 1$, implying $|\lambda| < 1$ and asymptotic stability.

When $\epsilon = 0$ or $\alpha \to \infty$, stability requires $|a| < 1$, i.e., $|r(1 - 2x^*)| < 1$. Memory thus imposes the stricter condition $|r(1 - 2x^*)| < 1 - \epsilon\, \zeta(\alpha)$, reducing the stable parameter range. For the logistic map, this means the maximum stable $r < 3$ is further reduced. When the perturbation is switched on slightly, the roots vary analytically as
\begin{equation}
\lambda = a + \epsilon\, \mathrm{Li}_\alpha(a^{-1}) + \mathcal{O}(\epsilon^2),
\label{eq10}
\end{equation}
since $\mathrm{Li}_\alpha(z)$ is analytic for $|z| < 1$ and $\alpha > 1$. As $\alpha \to \infty$, the kernel $k^{-\alpha} \to \delta_{k1}$, so that $\mathrm{Li}_\alpha(z) \to z$, and
\begin{equation}
\lambda = a + \epsilon a = a(1 + \epsilon).
\label{eq11}
\end{equation}
This corresponds to an effective rescaling; the Feigenbaum period-doubling cascade emerges with universal $\delta \approx 4.669$, perturbatively recovering the rescaled Markovian map
\begin{equation}
x_{n+1} = a(1 + \epsilon)\, x_n (1 - x_n),
\label{eq12}
\end{equation}
with effective parameter $r' = r(1 + \epsilon)$. Thus, memory effects vanish continuously, justifying the recovery of classical universality. The re-normalization group fixed points are structurally stable.
\subsubsection{Case II: Long Range Memory case ($\alpha \leq 1$)}
In this case, $\zeta(\alpha) = \infty$, the bound diverges and there is no finite stability criterion, implying that the memory term dominates non-perturbatively. Fixed-point instability occurs when a root crosses $|\lambda| = 1$, primarily near $\lambda = 1$ (neutral stability, as in the classical case). Set $\lambda = e^{\mu}$ with $\mu \to 0^{+}$, i.e., approaching from the region $|\lambda| < 1$ and substituting in Equation~(\ref{eq7}) gives
\begin{equation}
e^{\mu} = a + \epsilon\, \mathrm{Li}_\alpha(e^{-\mu}).
\label{eq13}
\end{equation}
For small $\mu$, the polylogarithm with $\alpha < 1$ has the singular asymptotic expansion near $z = 1$:
\begin{equation}
\mathrm{Li}_\alpha(e^{-\mu}) \approx \Gamma(1-\alpha)\, \mu^{\alpha-1},
\label{eq14}
\end{equation}
obtained from the series $\sum_{k=1}^\infty k^{-\alpha} e^{-k\mu}$ by matching the Mellin transform of the geometric series, with $\Gamma(1-\alpha)$ arising from the analytic continuation of $\mathrm{Li}_\alpha(z)$. Assuming the leading-order balance $e^\mu \approx 1 + \mu$ and $a \approx 1$, the characteristic equation becomes
\begin{equation}
1 + \mu \approx 1 + \epsilon\, \Gamma(1-\alpha)\, \mu^{\alpha-1},
\label{eq15}
\end{equation}
so that, to leading order,
\begin{equation}
\mu = \epsilon\, \Gamma(1-\alpha)\, \mu^{\alpha-1}.
\label{eq16}
\end{equation}

Assume $\mu \approx c\, \epsilon^\beta$ with $c > 0$, $\beta > 0$, and substitute into equation~(\ref{eq16}):
\begin{equation*}
c\, \epsilon^\beta = \epsilon\, \Gamma(1-\alpha)\, c^{\alpha-1} \epsilon^{\beta(\alpha-1)}.
\end{equation*}
Equating the exponents of $\epsilon$ on both sides,
\begin{equation}
\beta = 1 + \beta(\alpha - 1),
\label{eq17}
\end{equation}
which gives
\begin{equation}
\beta = \frac{1}{2 - \alpha}.
\label{eq18}
\end{equation}
The prefactor then satisfies
\begin{equation*}
c = \bigl[ \Gamma(1-\alpha)\, c^{\alpha-1} \bigr]^{1/(2-\alpha)},
\end{equation*}
which has a positive real solution for $c$. Thus,
\begin{equation}
\mu \sim \epsilon^{1/(2-\alpha)}.
\label{eq19}
\end{equation}
Since $\mu > 0$, $\lambda = e^{\mu} > 1$, regardless of how negative $a < 0$ is; the perturbative shift cannot compensate for the singular term $\mu^{\alpha-1}$. The fixed point is therefore unstable in a non-perturbative manner.
\subsubsection{Case III: Limiting Case ($\alpha=1$)}
At $\alpha = 1$, the polylogarithm reduces to the ordinary logarithm and long-range memory is marginally divergent:
\begin{equation}
\mathrm{Li}_1(e^{-\mu}) = -\ln(1 - e^{-\mu}) \approx -\ln \mu.
\label{eq20}
\end{equation}
Therefore, near $a \approx 1$,
\begin{equation}
\mu \approx -\epsilon \ln \mu,
\label{eq21}
\end{equation}
a transcendental equation for~$\mu$ in terms of~$\epsilon$. The asymptotic behaviour is
\begin{equation}
\mu \sim -\epsilon \ln \epsilon,
\label{eq22}
\end{equation}
showing that there is still no stability threshold: even a small $\epsilon$ destabilizes the fixed point.

Therefore, in summary, for $\alpha \leq 1$ and $\epsilon > 0$, the fixed points are generally unstable. The transition at $\alpha = 1$ is analytically sharp, distinguishing perturbative shrinkage of the stable regime ($\alpha > 1$) from absolute destabilization ($\alpha \leq 1$). This motivates a universality classification of chaos in non-Markovian dynamics, as summarized in the table below. $\alpha=1$ is identified as a critical dimension in time.

\vspace{0.2cm}
\begin{tabularx}{0.45\textwidth}{|>{\centering\arraybackslash}X|>{\centering\arraybackslash}X|>{\centering\arraybackslash}X|}
\hline
Memory exponent & Stability & Universality \\
\hline
$\alpha > 1$ & Perturbative shrinkage & Markovian (Feigenbaum) \\
\hline
$\alpha = 1$ & Marginal destabilization & Logarithmic \\
\hline
$\alpha < 1$ & Non-perturbative instability & Power-law \\
\hline
\end{tabularx}
\vspace{0.2cm}

\begin{figure}[t]
  \centering
  \includegraphics[width=0.9\columnwidth]{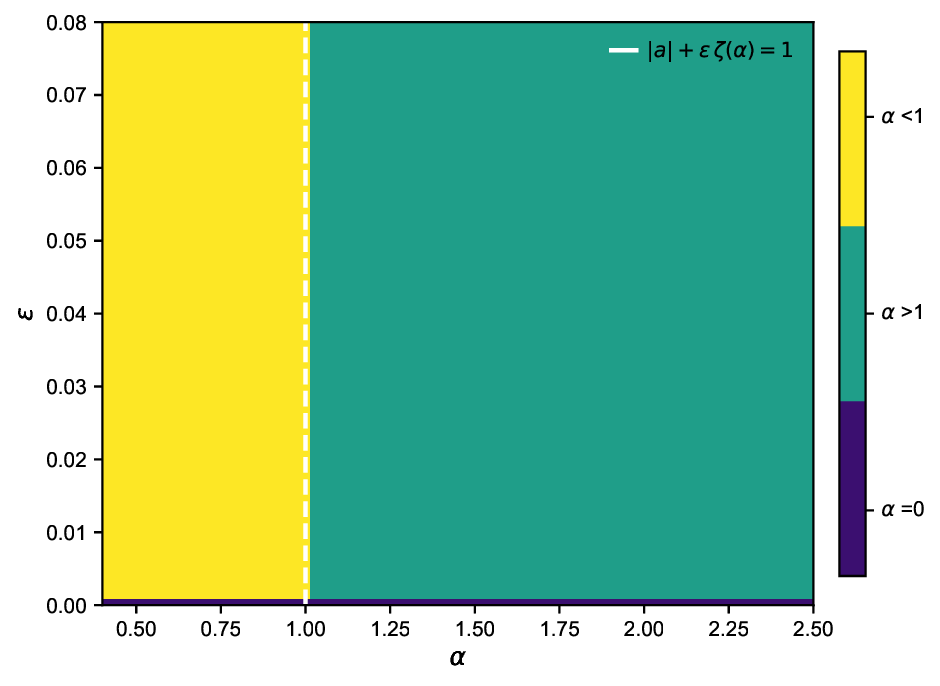}
  \caption{
    \textbf{Phase diagram of chaos universality classes in the non-Markovian logistic map.}
    The three regions correspond to: classical universality (period-doubling cascade),     deformed universality (memory-renormalized cascade), and memory-dominated chaos (fractional scaling, no Feigenbaum cascade). The dashed line at $\alpha=1$ separates     summable (short-range) and non-summable (long-range) memory. The solid white curve shows the analytic stability boundary $|a| + \epsilon\,\zeta(\alpha) = 1$ for $\alpha > 1$. Parameters: $r = 3.2$, $x_0 = 0.2$, $n_{\mathrm{iter}} = 8000$, $n_{\mathrm{transient}} = 5000$, memory cutoff $= 2000$.
  }
  \label{fig2}
\end{figure}

FIG.~\ref{fig2} summarizes the stability and universality structure in the memory--strength phase plane, identifying $\alpha = 1$ as a sharp boundary that separates perturbative (summable) memory from genuinely non-Markovian dynamics.
\section{Fractional Lyapunov Scaling}
Fractional Lyapunov scaling arises from memory-induced critical slowing down near the stability boundary, which leads to a non-classical exponent $\beta(\alpha) \ne 1/2$. The finite-time Lyapunov exponent measures the average logarithmic expansion rate:
\begin{equation}
\lambda_N = \frac{1}{N} \sum_{n=1}^{N} \ln \left| \frac{\partial x_{n+1}}{\partial x_n} \right|.
\label{eq23}
\end{equation}

For a non-Markovian map with power-law memory, the local expansion factor takes the form
\begin{equation}
\left| \frac{\partial x_{n+1}}{\partial x_n} \right| = \left| f'(x_n) + \epsilon \sum_{k=1}^{n} \frac{1}{k^\alpha} \, \frac{\partial x_{n-k}}{\partial x_n} \right|.
\label{eq24}
\end{equation}

Near a fixed point $x^*$, the dynamics can be linearized. In terms of the linearized modes, this becomes approximately
\begin{equation}
\left| a + \epsilon \sum_{k=1}^{n} k^{-\alpha} \lambda^{n-k} \right|,
\label{eq25}
\end{equation}
where $a = f'(x^*)$ is the local slope at the fixed point and $\lambda$ is the dominant eigenvalue (Lyapunov exponent) of the linearized system.
\subsection{Classical Case ($\alpha \rightarrow \infty$, Markovian)}
In this classical, short-memory limit, near a pitchfork bifurcation the relaxation rate scales linearly with the distance from criticality,

\begin{equation}
\mu \sim |r - r_c|,
\label{eq26}
\end{equation}

which corresponds to standard critical slowing down, and as a result the Lyapunov exponent also scales linearly,

\begin{equation}
\lambda(r) \sim |r - r_c|.
\label{eq27}
\end{equation}

In contrast, near a tangent (saddle-node) bifurcation — such as the onset of chaos in the logistic map via intermittency — the Lyapunov exponent scales as

\begin{equation}
\lambda(r) \sim |r - r_c|^{1/2},
\label{eq28}
\end{equation}

a non-classical exponent arising from the power-law (logarithmic in discrete time) relaxation and long laminar phases. This $\beta = 1/2$ scaling in the finite-time Lyapunov exponent is a hallmark of the classical discrete map at the onset of chaos and serves as the reference (Markovian) behavior against which the anomalous $\beta(\alpha)$ in the fractional/long-memory case is compared.

\subsection{Case I: Summable Memory($\alpha >1$)}
From the case I bound, the stability edge occurs at
\begin{equation}
a_c(\epsilon,\alpha) = -(1 - \epsilon \zeta(\alpha)).
\label{eq29}
\end{equation}

Near this edge, the relaxation rate is still linear (perturbative), $\mu \sim |a - a_c|$, but the critical point is shifted by the polylogarithmic factor. The shifted critical parameter is given by
\begin{equation}
r_c(\alpha) = \frac{r_c^0}{1 - \epsilon \zeta(\alpha)},
\label{eq30}
\end{equation}
where $r_c^0$ is the critical value in the Markovian ($\alpha \to \infty$) limit. With this rescaled critical point, the finite-time Lyapunov exponent still exhibits the square-root scaling
\begin{equation}
\lambda_N \sim \big( r - r_c(\alpha) \big)^{1/2},
\label{eq31}
\end{equation}
so that the critical exponent remains $\beta(\alpha) = 1/2$, but now measured with respect to the memory-dependent critical point $r_c(\alpha)$.
\subsection{Case II: Non Summable Memory($\alpha \leq 1$)}
From Case II, the marginal root of the fractional characteristic equation may be written as
$\lambda = e^{\mu}$, with $\mu \ll 1$. The instability threshold $a_c(\epsilon,\alpha)$
(or equivalently $r_c(\epsilon,\alpha)$) is determined by the condition
\begin{equation}
1 \approx a_c + \epsilon\,\Gamma(1-\alpha)\,\mu_c^{\alpha-1},
\label{eq32}
\end{equation}
where $\mu_c$ denotes the critical fractional growth rate. Solving this relation
self-consistently yields $\mu_c \sim \epsilon^{1/(2-\alpha)}$, implying that the shift
of the critical control parameter scales as
\begin{equation}
r_c(\epsilon,\alpha) - r_c^0 \sim \epsilon^{1/(2-\alpha)}.
\label{eq33}
\end{equation}
In the vicinity of $r_c$, the dynamics are governed by the slowest fractional mode,
whose growth rate vanishes as
\begin{equation}
\mu \sim |r-r_c|^{1/(2-\alpha)},
\label{eq34}
\end{equation}
leading to a divergence of the correlation time $\tau \sim \mu^{-1}$.

The finite-time Lyapunov exponent may be expressed as a temporal average of the
logarithmic expansion rate,
\begin{equation}
\lambda_N = \frac{1}{N}\int_0^N \ln|\lambda(t)|\,dt.
\label{eq35}
\end{equation}
Near the instability threshold, the dominant contribution arises from the fractional
eigenmode, for which $\lambda(t)\sim e^{\mu}$ varies slowly over the correlation time,
so that $\ln|\lambda(t)|\sim\mu$. Substituting into the above expression therefore gives
\begin{equation}
\lambda_N \sim \mu \sim |r-r_c|^{1/(2-\alpha)}.
\label{eq36}
\end{equation}
Thus, the Lyapunov exponent exhibits non-classical critical scaling,
\begin{equation}
\lambda \sim |r-r_c|^{\beta(\alpha)}, \qquad
\beta(\alpha)=\frac{1}{2-\alpha},
\label{eq37}
\end{equation}
demonstrating memory-induced critical dynamics and a genuinely non-Markovian route to
chaos.

\begin{figure*}[t]
  \centering
  \includegraphics[width=0.7\textwidth]{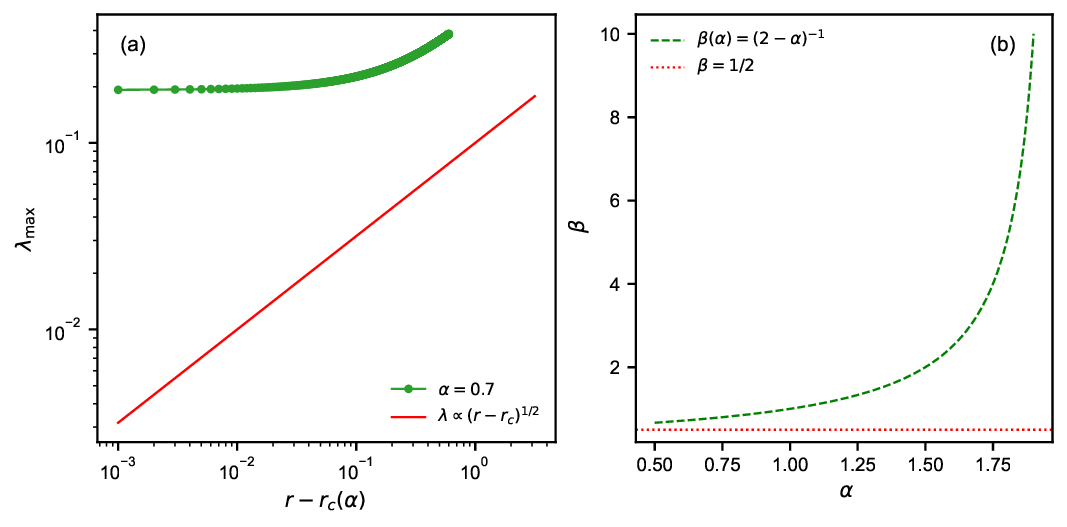}
  \caption{
    \textbf{Memory‑induced change in Lyapunov scaling.}
 (a) Lyapunov exponent~$\lambda_{\max}$ vs.~$r - r_c(\alpha)$ for the classical logistic map ($\alpha=\infty$), summable memory ($\alpha=1.5$), and long-range memory ($\alpha=0.7$). Classical and summable memory show the same square-root scaling~$\lambda \sim (r - r_c)^{1/2}$, while long-range memory exhibits a different, fractional scaling.
(b) Critical exponent~$\beta(\alpha)$ vs. memory exponent~$\alpha$. The numerical data (circles) are well described by~$\beta(\alpha) = (2-\alpha)^{-1}$ for long-range memory, while classical and summable memory fall on the universal value~$\beta = 1/2$.  
  }
  \label{fig3}
\end{figure*}

FIG~(\ref{fig3}) demonstrates that the Lyapunov exponent near the onset of chaos follows a fractional scaling law controlled by the memory exponent. For long-range memory ($\alpha = 0.7$), the numerical data systematically deviate from the classical square-root prediction and instead obey $\lambda \sim (r - r_c)^{1/(2 - \alpha)}$. Panel~(b) shows that this behavior is generic, with the Lyapunov critical exponent varying continuously with $\alpha$ and reducing to the classical value only in the Markovian limit. Thus, memory defines new universality classes of chaos.
\section{Implications and Outlook}
Universality is a cornerstone of chaos theory, epitomized by the Feigenbaum scenario and its remarkable insensitivity to microscopic details. This framework rests on the implicit Markovian assumption that the future evolution depends only on the present state, with no memory of the past. Our results show that this assumption has profound consequences: temporal correlations act as an independent control parameter that fundamentally reshapes the structure of chaotic dynamics, leading to a rich family of non-classical scaling laws beyond the standard universality classes.

We demonstrate, using a power-law memory kernel, that classical universality survives only when temporal correlations are summable. In this regime, the period-doubling cascade and the square-root Lyapunov scaling persist, as memory merely renormalizes the control parameters without altering the underlying bifurcation structure. However, long-range memory fundamentally changes the route to chaos, giving rise to fractional critical scaling of the Lyapunov exponent with a continuously tunable exponent and enabling chaos without a Feigenbaum cascade. Strong nonlinearity in this setting further suppresses chaotic fluctuations. The balance between local driving and history-dependent feedback is qualitatively reshaped by this non-Markovian mechanism, revealing a new class of critical behavior beyond standard universality.

Our results establish temporal correlations as a new axis of universality, analogous to spatial dimensionality in equilibrium critical phenomena. The memory exponent acts as an effective temporal dimension, with a sharp boundary separating perturbative (Markovian) and memory‑dominated regimes. Classical universality then emerges as a special limit within a broader universality landscape governed by the system’s temporal history.

Memory effects are so prevalent in real dynamical systems, such as in neural systems, climate models, viscoelastic materials, population dynamics, and active matter, due to slow relaxation, feedback, or collective interactions. Irregular dynamics can appear and disappear in such systems in ways inaccessible to Markovian models. The chaotic transitions may not follow the classical scaling laws. The fractional scaling of the Lyapunov exponent suggests that the chaotic fluctuations could serve as sensitive probes of hidden temporal correlations in empirical time series.

Several open theoretical questions arise naturally. The role of long‑range memory in reshaping universality in higher‑dimensional systems and continuous‑time flows remains largely unexplored, as does the formulation of a systematic renormalisation‑group theory for non‑Markovian dynamics in functional phase space. An important direction for future work is to investigate how memory‑induced effective dimensionality affects Lyapunov spectra, entropy production, and the geometry of attractors. A unified picture of memory‑driven non‑equilibrium dynamics emerges naturally by extending these ideas to stochastic systems with temporally correlated noise.

\section{Conclusion}
In summary, it has been demonstrated that long-range temporal correlations significantly alter the universality of chaos. Power-law memory breaks classical universality and gives rise to fractional critical behaviour, revealing that universality in dynamical systems is richer and a fragile domain of exploration when history dependence is unavoidable. This research opens new directions for exploring non-Markovian dynamics across physics, biology, and the Earth sciences.

\bibliography{bibliography}
\end{document}